\documentclass[aps,prl,twocolumn,showpacs,groupedaddress]{revtex4}

\usepackage{graphicx}
\usepackage{color}
\usepackage{amsmath}
\usepackage{hyperref}

\begin{document}

% Use the \preprint command to place your local institutional report
% number in the upper righthand corner of the title page in preprint mode.
% Multiple \preprint commands are allowed.
% Use the 'preprintnumbers' class option to override journal defaults
% to display numbers if necessary
%\preprint{}

%Title of paper
\title{Warm turbulence in the Boltzmann equation}
%\title{Turbulent cascades in the \blue{isotropic} \red{[PIETRO: maybe it is better to be precise]} Boltzmann kinetic equation}

% repeat the \author .. \affiliation  etc. as needed
% \email, \thanks, \homepage, \altaffiliation all apply to the current
% author. Explanatory text should go in the []'s, actual e-mail
% address or url should go in the {}'s for \email and \homepage.
% Please use the appropriate macro foreach each type of information

% \affiliation command applies to all authors since the last
% \affiliation command. The \affiliation command should follow the
% other information
% \affiliation can be followed by \email, \homepage, \thanks as well.

\author{Davide Proment}
\email{davideproment@gmail.com}
\homepage{http://www.to.infn.it/~proment}
\affiliation{Dipartimento di Fisica Generale, Universit\`{a} degli Studi di Torino, Via Pietro Giuria 1, 10125 Torino, Italy}
\affiliation{INFN, Sezione di Torino, Via Pietro Giuria 1, 10125 Torino, Italy}

\author{Sergey Nazarenko}
\affiliation{Mathematics Institute, The University of Warwick, Coventry, CV4-7AL, UK}

\author{Pietro Asinari}
%\homepage{http://www.institution.edu/~Author}
\affiliation{Dipartimento di Energetica, Politecnico di Torino, Corso Duca degli Abruzzi 24, 10129 Torino, Italy}

\author{Miguel Onorato}
\affiliation{Dipartimento di Fisica Generale, Universit\`{a} degli Studi di Torino, Via Pietro Giuria 1, 10125 Torino, Italy}
\affiliation{INFN, Sezione di Torino, Via Pietro Giuria 1, 10125 Torino, Italy}

%\email[]{Your e-mail address}
%\homepage[]{Your web page}
%\thanks{}
%\altaffiliation{}
%\affiliation{}

%Collaboration name if desired (requires use of superscriptaddress
%option in \documentclass). \noaffiliation is required (may also be
%used with the \author command).
%\collaboration can be followed by \email, \homepage, \thanks as well.
%\collaboration{}
%\noaffiliation

\date{\today}

\begin{abstract}
We study the single-particle distributions of three-dimensional hard sphere gas described by the Boltzmann equation.
We focus on the steady homogeneous isotropic solutions in thermodynamically open conditions, i.e. in the presence of forcing and dissipation. 
We observe nonequilibrium steady state solution characterized by a warm turbulence, that is an energy and particle cascade superimposed on the Maxwell-Boltzmann distribution. 
We use a dimensional analysis approach to relate the thermodynamic quantities of the steady state with the characteristics of the forcing and dissipation terms.
In particular, we present an analytical prediction for the temperature of the system which we show to be dependent only on the forcing and dissipative scales. 
Numerical simulations of the Boltzmann equation support our analytical predictions.
\end{abstract}

% insert suggested PACS numbers in braces on next line
% insert suggested keywords - APS authors don't need to do this
%\keywords{}

%\maketitle must follow title, authors, abstract, \pacs, and \keywords
\maketitle

% body of paper here - Use proper section commands
% References should be done using the \cite, \ref, and \label commands
% \section{}
% Put \label in argument of \section for cross-referencing
%\section{\label{}}
% \subsection{}
% \subsubsection{}

Three-dimensional fluid turbulence is characterized by an energy cascade process whose concept was introduced by Richardson in 1922 and developed by Kolmogorov in 1941 \cite{kolmogorov41, frisch1995t}. 
The idea is that if one injects energy at some scale, creating eddies, then those eddies may interact and create smaller and smaller eddies up to the dissipation scale. 
The result is  the famous power-law Kolmogorov  spectrum  which corresponds to a constant flux of energy from large to small scales.

Some years later, in 1965 \cite{zakharov1965}, Zakharov found that dispersive weakly nonlinear wave systems, e.g. surface gravity waves, may exhibit a very similar phenomenology as fluid turbulence, i.e. a constant flux of energy towards small scales, a {\it direct cascade}. 
Besides ocean waves \cite{zakharov1967ess}, these cascades have been studied in a large number of weakly nonlinear systems such as internal waves \cite{lvov2004eso}, nonlinear optics \cite{dyachenko:1992hc}, Bose-Einstein condensation \cite{PhysRevLett.74.3093, nazarenko2006wta, proment:051603}, magnetohydrodynamics \cite{galtier2000wtt}. 
Such regime, valid for weak nonlinearity, is now known as {\it weak  wave turbulence} and the power-law flux carrying states are called Kolmogorov-Zakharov (KZ) spectra.
The extraordinary fact is that the KZ solutions corresponding to the turbulent cascades are found to be exact analytical solutions of a {\it wave kinetic equation} which describes the evolution  of the wave spectrum. 
The wave kinetic equation has a structure that resembles the classical Boltzmann equation: the evolution of the distribution is driven by a collision integral which conserves mass, momentum and energy.

%We emphasize that in wave turbulence, the KZ solutions are relevant if one considers the system as thermodynamically open: a source (forcing) and a sink (dissipation) are localized in the momentum space. 
%The power-law distributions then represent the stationary nonequilibrium states. 
%In closed system, on the contrary, the wave kinetic equation has a thermodynamic equilibrium solution just like the Boltzmann equation and an {\it H-theorem} can be formulated \cite{zakharov41kst}.

Keeping in mind such an analogy, our aim is to address in the present Letter the following fundamental questions.
Let us consider a gas whose single-particle distribution evolution is described by the Boltzmann equation and suppose that we inject in the system particles with a specific energy and remove those that reach an energy larger/smaller than a preselected threshold.
How does the distribution function change with respect to the thermodynamic equilibrium solution? 
%Will we observe a constant flux of conserved quantities across some part of the distribution function?
%Do we expect to observe power-law solutions as in wave turbulence? %%%constant flux is a trivial consequence of definition and stationarity
How the temperature and chemical potential will be affected by the presence of forcing and dissipation? 
%Indeed, in statistical mechanics, investigating the general properties of a system in contact with reservoirs, namely an open system, is still a long lasting problem. 
% (e.g. see the second problem discussed by E.H. Lieb on the occasion of the award of the Boltzmann medal \cite{Lieb1999491}).

In order to answer we will first consider numerical simulations of the Boltzmann equation in the homogeneous and isotropic case and then present an argument based on dimensional analysis that allows us to explain the numerical results and make some predictions on the steady nonequilibrium properties of the system.
Our starting point is the homogeneous, forced and damped Boltzmann equation
\begin{equation}
\begin{split}
& \frac{\partial n_1}{\partial t}   = I_{coll} + F - D, \, \, \, \, \mbox{with} \\
& I_{coll}=\int_{-\infty}^{+\infty} W_{12}^{1'2'} \left[n'_1n'_2-n_1n_2\right] d\mathbf{v}_2 d\mathbf{v}^{\prime}_1 d\mathbf{v}^{\prime}_2,
\end{split}
\label{eq:BE}
\end{equation}
%with
%\begin{equation}
%I_{coll}=\int_{-\infty}^{+\infty} W_{12}^{1'2'} \left[n'_1n'_2-n_1n_2\right] d\mathbf{v}_2 d\mathbf{v}^{\prime}_1 d\mathbf{v}^{\prime}_2
%\end{equation}
where  $ n_i \equiv n({\bf x},{\bf v}_i,t ) $ is the single-particle distribution function and primes denote particles after the collision. 
We have included a source term $ F $ and a sink $ D $ which will be specified once the numerics are discussed. 
As we consider elastic collisions, the general way to express $ W $ is
\begin{equation}
W_{12}^{1'2'}=\sigma \, \delta(\mathbf{v}_1+\mathbf{v}_2-\mathbf{v}^{\prime}_1-\mathbf{v}^{\prime}_2)  \, \delta(E_1+E_2 - E_1^\prime-E_2^\prime),
\end{equation}
with $E_i=|{\bf v}_i|^2 /2 $  being the kinetic energy per unit mass.
The $ \delta $-functions assure conservation of the total momentum and the total kinetic energy. 
In this Letter, we consider the three-dimensional rigid sphere gas ($ \sigma$ is independent of ${\bf v}$) in isotropic conditions. 
%Therefore, before performing numerical computations, we write the equation (\ref{eq:BE}) in the energy domain, $ E $-space, by consider spherical coordinates and integrate over solid angles.

In absence of forcing and dissipation any initial condition will relax to the Maxwell-Boltzmann (MB) distribution
\begin{equation}
n_{MB}(E)= {n_0}\,e^{-\frac{E+\mu}{T}}= A \, e^{-\frac{E}{T}},
\label{eq:MB}
\end{equation}
where $ A\equiv{n_0}\,e^{-\frac{\mu}{T}}$ with  $\mu$ and $T$  the chemical potential and temperature of the system respectively.
This equipartition mechanism, consequence of the H-theorem, has been checked as a benchmark of our numerical code. 
In order to consider an open system we have then included forcing and dissipation written in $ E $-space. 
The forcing term 
%$ \mathcal{F}= 2\pi \int F(E) \, E^{1/2} dE $ 
is constant in time and has the role of injecting particles with energies narrowly concentrated around some value  $ E_f $,
i.e. $F(E)  = F $ if $|E-E_f|< \delta_f$ and zero otherwise,  where $ F $ is a positive constant.
The dissipation term
%$ \mathcal{D} = 2\pi \int D(E) \, E^{1/2} dE $ 
is implemented as a filter which removes,  at each iteration time, energy and particles outside of the domain $ \left(E_{\min}, E_{\max} \right) $,
i.e. $D(E) = 0$  if $ E \in \left(E_{\min}, E_{\max} \right) $ and $D(E) = - \infty$ otherwise. 
In such conditions, a dimensional analysis shows that 
 (\ref{eq:BE}) may have also solutions characterized by 
a  constant flux of energy and mass which corresponds to 
the KZ solutions. Indeed,
the particle flux $\eta$ and the energy flux $\epsilon$ can be estimated as:
\begin{equation}
\begin{split}
& \eta = 2\pi\int_0^E \frac{\partial n}{d t} \, E^{1/2} dE  \sim n^2 E^{7/2} \\
& \epsilon = 2\pi\int_0^E \frac{\partial n}{d t} \, E^{3/2} dE  \sim n^2 E^{9/2}.
\label{eq:flux}
\end{split}
\end{equation}
Assuming that one of the fluxes is constant trough energy scales,  we can immediately derive the KZ solutions $ n_\eta \sim E^{-7/4} $ and $ n_\epsilon \sim E^{-9/4} $ \cite{kats1973symmetry}.

\begin{figure}
\includegraphics[scale=1.3]{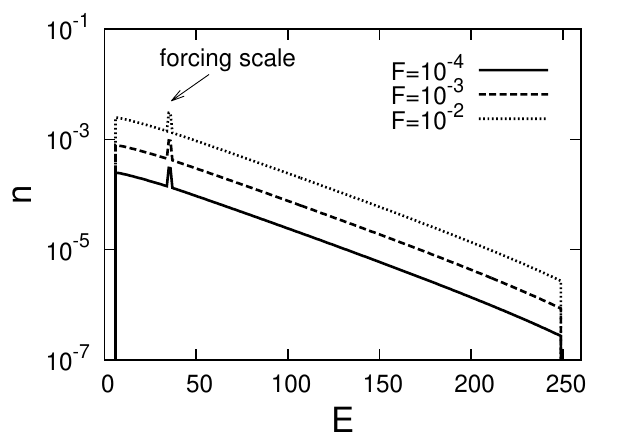}
\caption{Final steady states of the Boltzmann equation (\ref{eq:BE}) obtained for different values of the forcing amplitude $ F $.
Axes are in lin-log coordinates. 
$ E_{\min}=5 $, $ E_{\max}=250 $ and $ E_f=35-37 $. \label{fig:exampleSpectra}}
\end{figure}

\begin{figure}
\includegraphics[scale=1.3]{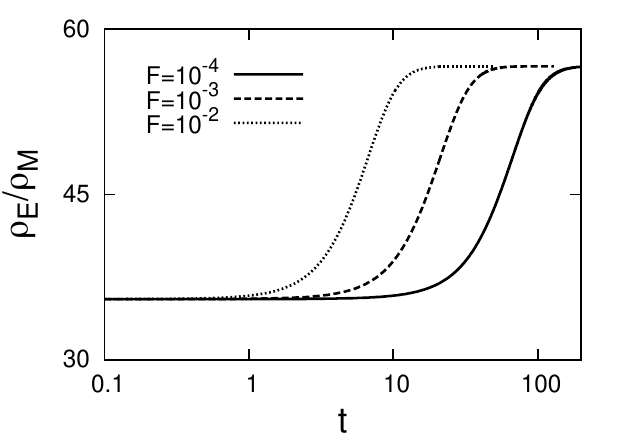}
\caption{Time evolution of the average energy per particle for different forcing amplitudes.
Axes are in log-lin coordinates.
$ E_{\min}=5 $, $ E_{\max}=250 $ and $ E_f=35-37 $. \label{fig:exampleEnergyMassRatio}}
\end{figure}

In our simulations the computational domain is uniformly discretized in 501 points having $ \Delta E= 1 $ and the initial  distribution is $ n(E, t=0)=0 $. 
We first consider the case characterized by $ E_{\min}=5 $, $ E_{\max}=250 $ and the forcing located at energies between 35 and 37.
Numerical results for the final steady states evaluated for three different forcing amplitudes,  $ F=10^{-4}, 10^{-5}, 10^{-6} $, are presented in Fig. \ref{fig:exampleSpectra}.
Independently of the forcing, a stationary distribution is reached in the simulations, implying the existence of a flux from the forcing region to the boundaries of the domain.
%In a close system the curves would approach the Maxwell-Boltzmann distribution which, in the log-lin coordinate, is represented by a straight line.  
The steady solutions of our simulation do not appear to be very far from the Maxwell-Boltzmann distribution (\ref{eq:MB}), which corresponds to a straight line in a lin-log coordinates. 
It is clear from the plot that the only consequence of increasing the forcing rate is to shift upwards the curves, leaving unchanged 
the slopes. This is the first indication that the temperature of 
the system remains constant, independently of the forcing.
%In order to be more precise we compute the temperature and the chemical potential from the Maxwell-Boltzmann distribution for the three cases.
We can also compute, as a function of time,  the average energy per particle $ \rho_E/\rho_M $, with $ \rho_M = 2\pi \int n(E) \, E^{1/2} \, dE $ and $ \rho_E =2\pi \int n(E) \, E^{3/2} \, dE $. 
As shown in Fig. \ref{fig:exampleEnergyMassRatio}, for large times  this quantity reaches a unique value for the three simulations considered. We recall that for a pure MB distribution such ratio is 
proportional to 3/2 the temperature of the system.
%
%{\color{red} [INSET???]} 

We have not observed KZ solutions in our simulations.
The reason is that the interactions are non-local in scales, as already pointed out in \cite{kats1975}: the collision integral does not converge for such solutions.
Moreover, the energy and particle flux directions, \cite{kats1976}, associated to such solutions, are opposite to the 
one predicted by the Fj{\o}rtoft argument which imposes that the energy should have a direct cascade, i.e. from low to high energies, while particle an inverse one \cite{2011arXiv1101.4137P}. 

As shown in Fig. \ref{fig:exampleSpectra}, we observe distributions not far from the thermodynamic equilibrium (\ref{eq:MB}) solution.
However, we are in a forced and dissipated situation and let us assume the existence of a small but finite flux correction  ``living on top" of the Maxwell-Boltzmann distribution. This behavior, named  {\it warm cascade}, has already been observed in other physical systems \cite{dyachenko:1992hc, nazarenko2004warm} and is characterized by constant flux cascades perturbing the thermodynamic equilibrium distribution. 
Mathematically, we assume that 
\begin{equation}
 n = n_{MB} \, (1+\tilde{n}),
\end{equation}
with $\tilde{n}$ the deviations with respect to the MB distribution
which are responsible for the fluxes; note that  not necessarily $\tilde{n}$
is small with respect to one.   If we plug such {\it ansatz}
in the equations for the fluxes (\ref{eq:flux})
 we obtain
\begin{equation}
\begin{split}
& \eta=c_1 \, n_{MB}^2 \, \tilde{n} \, (2+\tilde{n}) \, E^{7/2} \\
& \epsilon = c_2 \, n_{MB}^2 \, \tilde{n} \, (2+\tilde n) \, E^{9/2},
\label{eq:flux'}
\end{split}
\end{equation}
where $c_1$ and $c_2$ are two constants which cannot be 
determined through dimensional analysis. 
The term $n_{MB}^2$, corresponding to the unperturbed MB distribution, does not give any contribution because it is not responsible for any net flux.
% because it reduces the collision integral to zero. %SN: kz also reduce it to 0, so?

\begin{figure}
\includegraphics[scale=1.3]{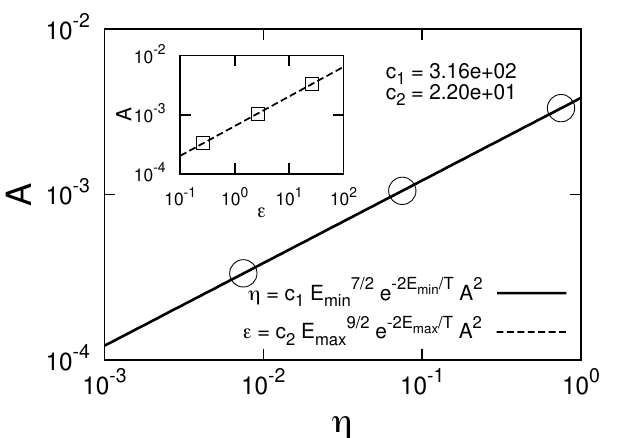}
\caption{$ A $ as a function of the fluxes. Circles (and squares in the inset) correspond to the numerical simulation results. 
The continuous and dashed lines are the predictions (\ref{eq:pred-flux}) where the temperature, $ T = 2 \rho_E / 3 \rho_M $, has been defined using the pure MB distribution. 
The values of the constant $ c_1 $ and $ c_2 $, shown in the plot, are found by a fit. 
\label{fig:AOfEtaAndEpsilon}}
\end{figure}

Our aim is to relate the macroscopic  properties of the system with the forcing and dissipation rates.
From our numerical computation we observe that, as we get closer to the  
cut-off scales ($ E_{\min} $ and $ E_{\max} $), deviations from a 
pure MB distribution becomes more relevant. 
Consequently, $ \tilde{n} $ becomes of the order one for $E=E_{\min}$ or $E=E_{\max}$ and therefore from (\ref{eq:flux'}) we have
\begin{equation}
\begin{split}
&   \eta = c_1 \, A^2 \, e^{-\frac{2 \,E_{\min}}{T}} \, E_{\min}^{7/2}  \\
&  \epsilon = c_2 \, A^2 \, e^{-\frac{2 \,E_{\max}}{T}} \, E_{\max}^{9/2},
\label{eq:pred-flux}
\end{split}
\end{equation}
where we have redefined the constant $c_1$ and $c_2$.
We now  verify the above relations through the direct computation of the 
Boltzmann equation. In Fig. \ref{fig:AOfEtaAndEpsilon} we show the dependence of  $ A $ on the fluxes  for three simulations previously described (Fig. \ref{fig:exampleSpectra}).
Supposing to use as temperature of the system the simple relation coming from the MB distribution $ T = 2\rho_E /3\rho_M $, we can observe in Fig. \ref{fig:AOfEtaAndEpsilon} (and its inset) that $ A $ scales as the square root of the incoming fluxes for fixed dissipative scales $ E_{\min} $ and $ E_{\max} $. 
From the numerics we can estimate, by a fit, the constants $ c_1 $ and $ c_2 $ whose values are reported in the figure.

We are now able to predict the dependence of the temperature on the forcing and the dissipative scales. Assuming that they are widely separated, that is $ E_{\min} \ll E_f \ll E_{\max} $, we have $ \epsilon = \eta \, E_f $. 
Then from equations (\ref{eq:pred-flux}) we get
\begin{equation}
T = \frac{2 \, (E_{\max}-E_{\min})}{\frac{9}{2}\ln{E_{\max}} - \frac{7}{2}\ln{E_{\min}} -\ln{E_f} + c_3 }
\label{eq:T}
\end{equation}
with $ c_3 = \ln{(c_2/c_1)} $. 
The temperature of the system does not depend on the incoming fluxes but only on the forcing and dissipative scales; this is consistent with results of the numerical simulations presented in Fig. \ref{fig:exampleSpectra} and \ref{fig:exampleEnergyMassRatio}. 

We can check the validity of prediction (\ref{eq:T}) by considering different simulations of the isotropic Boltzmann equation (\ref{eq:BE}) changing the forcing and dissipative scales. 
Before entering in the details, we emphasize that our 
predictions are based on a dimensional argument and  $ c_1 $ and $ c_2 $, which define $ c_3 $, can only be measured {\it via} numerical computations. 
In our simulations we have observed that, as $ E_{\min} $, $ E_f $ and $ E_{\max} $ changes, $ c_1 $ and $ c_2 $ assumes different values.
This may be related to the fact that we are considering the hypothesis that the particle flux is all dissipated at low energy scales and all energy flux at high ones, that is $ E_{\min} \ll E_f \ll E_{\max} $. 
This is not always verified in the numerical simulations due to finiteness of the computational domain.
In our numerical simulations we  have measured  an upper and 
lower bound for our constants leading to  $-4.21 \le c_3\le-1.05 $. 
This interval includes the analytical prediction $ c_3 = 2 \ln{(2/9)} $ obtained using a {\it diffusion approximation model} in the limit $ E_{\min} \ll T \ll E_{\max} $ presented in \cite{2011arXiv1101.4137P}.

\begin{figure}
\includegraphics[scale=1.3]{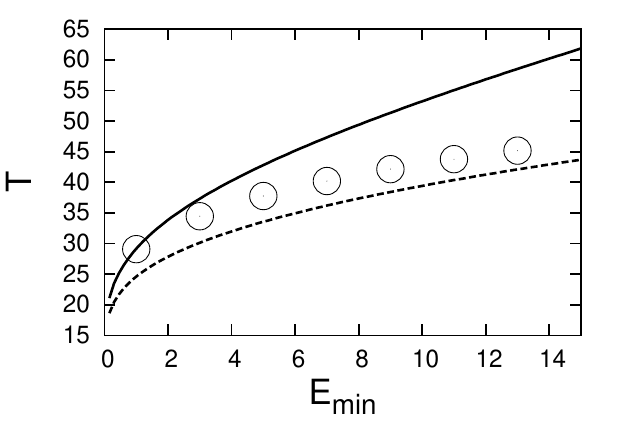}
\caption{Temperature as a function of the low energy cut-off $ E_{\min} $, keeping fixed $ F=10^{-3} $, $ E_f=36 $, $\delta_f=1$ and $ E_{\max}=250 $. 
Circles represent the results from numerical computations. 
The continuous and dashed lines are the prediction (\ref{eq:T}) for different values of $ c_3 $. 
\label{fig:temperatureOfEmin}}
\end{figure}

In Fig. \ref{fig:temperatureOfEmin} we show the comparison between  the estimation of the temperature  from (\ref{eq:T}) and the numerical simulation varying the low energy dissipation scale $ E_{\min}$. 
The incoming flux, the forcing and the high energy dissipative scale have been kept fixed to the respective values of $ F=10^{-3} $, $ E_f=36 $, $\delta_f=1$, $ E_{\max}=250 $.
The dashed and the solid line in the figure are the predictions obtained with  $c_3=-1.05$ and $c_3=-4.21$, respectively.  
The agreement between the temperature prediction and the computation is satisfactory. 

\begin{figure}
\includegraphics[scale=1.3]{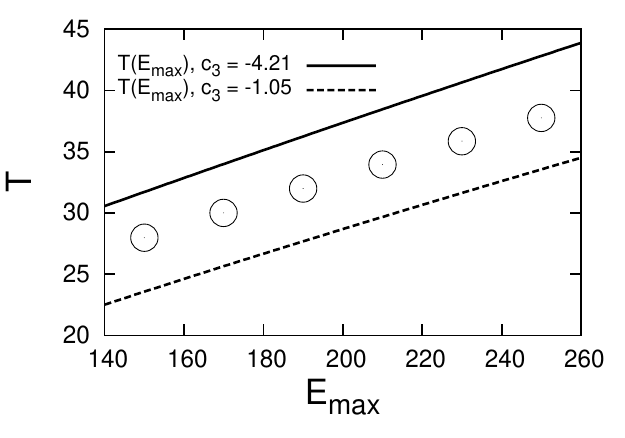}
\caption{Temperature as a function of the high energy cut-off $ E_{\max} $, keeping fixed $ F=10^{-3} $, $ E_{\min}=5 $ and $ E_f=36 $, $\delta_f=1$.
Circles represent the results from numerical computations. 
The continuous and dashed lines are the prediction (\ref{eq:T}) for different values of $ c_3 $. 
\label{fig:temperatureOfEmax}}
\end{figure}

In Fig. \ref{fig:temperatureOfEmax} we present the behavior of  the temperature as a function of  the high energy cut-off scale $ E_{\max} $, with $ F=10^{-3} $, $ E_{\min}=5 $ and $ E_f=36 $, $\delta_f=1$.
The comparison between our estimation  (\ref{eq:T}) and the numerics  is very good.

In the present Letter we have investigated the stationary states of a three-dimensional hard sphere gas whose single-particle distribution function is modeled by the homogeneous isotropic Boltzmann equation.
In particular, we were interested in the steady nonequilibrium states in an open system condition, i.e. in the presence of a forcing and dissipation mechanisms. 
Using the language of wave turbulence theory we have assumed the presence of a {\it warm cascades} as steady distributions. 
These solutions are characterized by constant particle and energy fluxes superimposed on the thermodynamic Maxwell-Boltzmann distribution.
Our assumption has allowed us  to relate the thermodynamic quantities of the system to the characteristics of the forcing and dissipative terms of the equation.

In particular, we have been able to give an analytical relation for the temperature reached in the system as a function of the forcing and 
dissipative scales.  
One of the main results is that the temperature is  independent of how much energy and particles we inject in the system but depends only on the cut-off scales and on the forcing scale.
By a direct numerical integration of the Boltzmann equation we have shown that our numerical results are consistent with the aforementioned prediction.

We believe that the  approach used in this Letter may open some perspectives towards understanding nonequilibrium steady states in other physical systems via analogies to cases widely studied by the wave turbulence theory.

% Create the reference section using BibTeX:
\bibliography{references}

\begin{thebibliography}{15}
\expandafter\ifx\csname natexlab\endcsname\relax\def\natexlab#1{#1}\fi
\expandafter\ifx\csname bibnamefont\endcsname\relax
  \def\bibnamefont#1{#1}\fi
\expandafter\ifx\csname bibfnamefont\endcsname\relax
  \def\bibfnamefont#1{#1}\fi
\expandafter\ifx\csname citenamefont\endcsname\relax
  \def\citenamefont#1{#1}\fi
\expandafter\ifx\csname url\endcsname\relax
  \def\url#1{\texttt{#1}}\fi
\expandafter\ifx\csname urlprefix\endcsname\relax\def\urlprefix{URL }\fi
\providecommand{\bibinfo}[2]{#2}
\providecommand{\eprint}[2][]{\url{#2}}

\bibitem[{\citenamefont{Kolmogorov}(1941)}]{kolmogorov41}
\bibinfo{author}{\bibfnamefont{A.}~\bibnamefont{Kolmogorov}}, in
  \emph{\bibinfo{booktitle}{Proceedings (Doklady) Academy of Sciences, USSR}}
  (\bibinfo{year}{1941}), vol.~\bibinfo{volume}{30}, pp.
  \bibinfo{pages}{301--305}.

\bibitem[{\citenamefont{Frisch}(1995)}]{frisch1995t}
\bibinfo{author}{\bibfnamefont{U.}~\bibnamefont{Frisch}},
  \emph{\bibinfo{title}{{Turbulence: the legacy of AN Kolmogorov}}}
  (\bibinfo{publisher}{Cambridge University Press}, \bibinfo{year}{1995}).

\bibitem[{\citenamefont{Zakharov}(1965)}]{zakharov1965}
\bibinfo{author}{\bibfnamefont{V.~E.} \bibnamefont{Zakharov}},
  \bibinfo{journal}{Journal of Applied Mechanics and Technical Physics}
  \textbf{\bibinfo{volume}{6}}, \bibinfo{pages}{22} (\bibinfo{year}{1965}),
  ISSN \bibinfo{issn}{0021-8944}, \bibinfo{note}{10.1007/BF01565814},
  \urlprefix\url{http://dx.doi.org/10.1007/BF01565814}.

\bibitem[{\citenamefont{Zakharov and Filonenko}(1967)}]{zakharov1967ess}
\bibinfo{author}{\bibfnamefont{V.}~\bibnamefont{Zakharov}} \bibnamefont{and}
  \bibinfo{author}{\bibfnamefont{N.}~\bibnamefont{Filonenko}}, in
  \emph{\bibinfo{booktitle}{Soviet Physics Doklady}} (\bibinfo{year}{1967}),
  vol.~\bibinfo{volume}{11}, p. \bibinfo{pages}{881}.

\bibitem[{\citenamefont{Lvov et~al.}(2004)\citenamefont{Lvov, Polzin, and
  Tabak}}]{lvov2004eso}
\bibinfo{author}{\bibfnamefont{Y.}~\bibnamefont{Lvov}},
  \bibinfo{author}{\bibfnamefont{K.}~\bibnamefont{Polzin}}, \bibnamefont{and}
  \bibinfo{author}{\bibfnamefont{E.}~\bibnamefont{Tabak}},
  \bibinfo{journal}{Physical Review Letters} \textbf{\bibinfo{volume}{92}},
  \bibinfo{pages}{128501} (\bibinfo{year}{2004}).

\bibitem[{\citenamefont{Dyachenko et~al.}(1992)\citenamefont{Dyachenko, Newell,
  Pushkarev, and Zakharov}}]{dyachenko:1992hc}
\bibinfo{author}{\bibfnamefont{S.}~\bibnamefont{Dyachenko}},
  \bibinfo{author}{\bibfnamefont{A.~C.} \bibnamefont{Newell}},
  \bibinfo{author}{\bibfnamefont{A.}~\bibnamefont{Pushkarev}},
  \bibnamefont{and} \bibinfo{author}{\bibfnamefont{V.~E.}
  \bibnamefont{Zakharov}}, \bibinfo{journal}{Physica D: Nonlinear Phenomena}
  \textbf{\bibinfo{volume}{57}}, \bibinfo{pages}{96} (\bibinfo{year}{1992}),
  \urlprefix\url{http://www.sciencedirect.com/science/article/B6TVK-46JH21H-4G%
/2/b9bf3a47086f6f154a8c0478ca64c07b}.

\bibitem[{\citenamefont{Semikoz and Tkachev}(1995)}]{PhysRevLett.74.3093}
\bibinfo{author}{\bibfnamefont{D.~V.} \bibnamefont{Semikoz}} \bibnamefont{and}
  \bibinfo{author}{\bibfnamefont{I.~I.} \bibnamefont{Tkachev}},
  \bibinfo{journal}{Phys. Rev. Lett.} \textbf{\bibinfo{volume}{74}},
  \bibinfo{pages}{3093} (\bibinfo{year}{1995}).

\bibitem[{\citenamefont{Nazarenko and Onorato}(2006)}]{nazarenko2006wta}
\bibinfo{author}{\bibfnamefont{S.}~\bibnamefont{Nazarenko}} \bibnamefont{and}
  \bibinfo{author}{\bibfnamefont{M.}~\bibnamefont{Onorato}},
  \bibinfo{journal}{Physica D: Nonlinear Phenomena}
  \textbf{\bibinfo{volume}{219}}, \bibinfo{pages}{1} (\bibinfo{year}{2006}).

\bibitem[{\citenamefont{Proment et~al.}(2009)\citenamefont{Proment, Nazarenko,
  and Onorato}}]{proment:051603}
\bibinfo{author}{\bibfnamefont{D.}~\bibnamefont{Proment}},
  \bibinfo{author}{\bibfnamefont{S.}~\bibnamefont{Nazarenko}},
  \bibnamefont{and} \bibinfo{author}{\bibfnamefont{M.}~\bibnamefont{Onorato}},
  \bibinfo{journal}{Physical Review A (Atomic, Molecular, and Optical Physics)}
  \textbf{\bibinfo{volume}{80}}, \bibinfo{eid}{051603}
  (pages~\bibinfo{numpages}{4}) (\bibinfo{year}{2009}),
  \urlprefix\url{http://link.aps.org/abstract/PRA/v80/e051603}.

\bibitem[{\citenamefont{Galtier et~al.}(2000)\citenamefont{Galtier, Nazarenko,
  Newell, and Pouquet}}]{galtier2000wtt}
\bibinfo{author}{\bibfnamefont{S.}~\bibnamefont{Galtier}},
  \bibinfo{author}{\bibfnamefont{S.}~\bibnamefont{Nazarenko}},
  \bibinfo{author}{\bibfnamefont{A.}~\bibnamefont{Newell}}, \bibnamefont{and}
  \bibinfo{author}{\bibfnamefont{A.}~\bibnamefont{Pouquet}},
  \bibinfo{journal}{Journal of Plasma Physics} \textbf{\bibinfo{volume}{63}},
  \bibinfo{pages}{447} (\bibinfo{year}{2000}).

\bibitem[{\citenamefont{Kats and Kontorovich}(1973)}]{kats1973symmetry}
\bibinfo{author}{\bibfnamefont{A.}~\bibnamefont{Kats}} \bibnamefont{and}
  \bibinfo{author}{\bibfnamefont{V.}~\bibnamefont{Kontorovich}},
  \bibinfo{journal}{Soviet Physics-JETP} \textbf{\bibinfo{volume}{37}},
  \bibinfo{pages}{80} (\bibinfo{year}{1973}).

\bibitem[{\citenamefont{Kats et~al.}(1975)\citenamefont{Kats, Kontorovich,
  Moiseev, and Novikov}}]{kats1975}
\bibinfo{author}{\bibfnamefont{A.}~\bibnamefont{Kats}},
  \bibinfo{author}{\bibfnamefont{V.}~\bibnamefont{Kontorovich}},
  \bibinfo{author}{\bibfnamefont{S.}~\bibnamefont{Moiseev}}, \bibnamefont{and}
  \bibinfo{author}{\bibfnamefont{V.}~\bibnamefont{Novikov}},
  \bibinfo{journal}{ZhETF Pis ma Redaktsiiu} \textbf{\bibinfo{volume}{21}},
  \bibinfo{pages}{13} (\bibinfo{year}{1975}).

\bibitem[{\citenamefont{Kats}(1976)}]{kats1976}
\bibinfo{author}{\bibfnamefont{A.}~\bibnamefont{Kats}},
  \bibinfo{journal}{Soviet Journal of Experimental and Theoretical Physics}
  \textbf{\bibinfo{volume}{44}}, \bibinfo{pages}{1106} (\bibinfo{year}{1976}).

\bibitem[{\citenamefont{{Proment} et~al.}(2011)\citenamefont{{Proment},
  {Onorato}, {Asinari}, and {Nazarenko}}}]{2011arXiv1101.4137P}
\bibinfo{author}{\bibfnamefont{D.}~\bibnamefont{{Proment}}},
  \bibinfo{author}{\bibfnamefont{M.}~\bibnamefont{{Onorato}}},
  \bibinfo{author}{\bibfnamefont{P.}~\bibnamefont{{Asinari}}},
  \bibnamefont{and}
  \bibinfo{author}{\bibfnamefont{S.}~\bibnamefont{{Nazarenko}}},
  \bibinfo{journal}{ArXiv e-prints}  (\bibinfo{year}{2011}),
  \eprint{1101.4137}.

\bibitem[{\citenamefont{Connaughton and Nazarenko}(2004)}]{nazarenko2004warm}
\bibinfo{author}{\bibfnamefont{C.}~\bibnamefont{Connaughton}} \bibnamefont{and}
  \bibinfo{author}{\bibfnamefont{S.}~\bibnamefont{Nazarenko}},
  \bibinfo{journal}{Phys. Rev. Lett.} \textbf{\bibinfo{volume}{92}},
  \bibinfo{pages}{044501} (\bibinfo{year}{2004}).

\end{thebibliography}

\end{document}